\documentclass[twocolumn,showpacs,prb,epsfig]{revtex4}
\usepackage{amsmath,amssymb}
\usepackage{graphicx,bm}
\usepackage{epsfig}
\newcommand{\Tr}{{\rm Tr}}

\begin{document}
\def\be{\begin{equation}}
\def\ee{\end{equation}}
\def\Tr{{\rm Tr}}
\title{Thermodynamics of ultra-small metallic grains in the auxiliary-field Monte Carlo approach}
\author{Y. Alhassid, L. Fang, and S. Schmidt}
\affiliation{Center for Theoretical Physics, Sloane Physics
  Laboratory, Yale University,  New Haven, Connecticut 06520, USA}
\begin{abstract}
We use an auxiliary-field Monte Carlo (AFMC) method to calculate thermodynamic properties (spin susceptibility and heat capacity) of ultra-small metallic grains in the presence of pairing correlations. This method allows us to study the crossover from bulk systems, where mean-field
BCS theory is valid, to the fluctuation-dominated regime of
ultra-small particles at finite temperature. The computational
effort at low temperatures is significantly reduced by exploiting a simple renormalization method.
\end{abstract}
\maketitle

\section{Introduction}
The properties of ultra-small metallic grains
(nanoparticles) differ from those of bulk
superconductors. Larger grains are well described by the Bardeen-Cooper-Scrieffer (BCS) mean-field theory~\cite{BCS}.
However, as the size of the grain becomes smaller, fluctuations around the BCS mean-field solution become important. This crossover from the BCS limit to the
fluctuation-dominated regime is controlled by the ratio $\Delta/\delta$ between the bulk BCS gap $\Delta$ and the mean level spacing $\delta$. BCS theory
is valid for $\Delta/\delta \gg 1$, while fluctuations become important as this ratio decreases and becomes similar to or smaller than $\sim 1$.~\cite{vonDelft}

Most of the recent interest in such ultra-small metallic grains has been
motivated by experiments
in the mid 1990s~\cite{RBT} in which the spectra of individual aluminum grains were determined by measuring the non-linear conductance. These spectra were found to depend on the magnetic field strength as well as the number parity of electrons
in the nanoparticle.  Pairing correlations were clearly observed in the larger metallic particles, and the results were explained qualitatively using a number-parity
projected BCS model.~\cite{tichy} However, for smaller particles with nominal radii in
the range $\sim 2-5$ nm it was not possible to distinguish between
the spacing due to the discreteness of the single-particle spectrum and the pairing
gap. The authors of Ref.~\onlinecite{RBT} emphasized that this does not mean that pairing
correlations cannot be observed in these samples but rather that
conductance measurements might not be the appropriate tool to study
the fluctuation-dominated regime.

It was suggested that number-parity effects of pairing correlations in the
fluctuation-dominated regime can be observed in
thermodynamic quantities such as the spin susceptibility.~\cite{lorenzo} Using a combination of different methods, the authors of Ref.~\onlinecite{lorenzo} found a reentrant behavior of
the spin susceptibility of grains with an odd number of electrons. For low temperatures they exploited an exact
solution of the BCS pairing Hamiltonian known as the Richardson
solution.~\cite{richardson} Results were presented for
$\Delta/\delta \le 1$ and $T \le 1.6\,\delta$ because the Richardson
solution becomes less tractable at larger values of $\Delta/\delta$
and/or higher temperatures. Their results at high temperatures were
based on the static-path approximation,~\cite{SPA} which ignores
quantum fluctuations. Signatures of pairing correlations in the spin susceptibility of ultra-small grains were also studied in Ref.~\onlinecite{schechter}.

Here we present a method to study thermal quantities such
as spin susceptibility and heat capacity in the full temperature
range and for arbitrary values of the ratio $\Delta/\delta$. We
use the auxiliary-field Monte Carlo (AFMC) method
developed in the framework of the interacting nuclear shell model, which is
 known (in nuclear physics) as the shell model Monte Carlo (SMMC) method.~\cite{smmc,smmc-app}
The SMMC approach was used to study thermal signatures of pairing correlations in the heat capacity~\cite{liu,level} and the moment of inertia~\cite{inertia} of
nuclei. Similar methods were used in the study of strongly correlated electron systems.~\cite{LG92} Here we adapt the AFMC method to study
pairing correlations in metallic nanoparticles. The dimension of the many-particle space depends combinatorially on the number of single-particle levels and/or number of electrons, and a direct diagonalization of the many-body Hamiltonian becomes impractical.
AFMC scales as a power law in the number of single-particle orbitals and can be carried out in very large model spaces. Recently, a different Monte Carlo approach that is based on canonical non-local loop update algorithm and is specific for a pairing Hamiltonian was used to study metallic nanoparticles.~\cite{vanhoucke}

The thermodynamic properties of grains with chaotic single-particle dynamics are expected to display mesoscopic fluctuations because of the randomicity associated with their single-particle Hamiltonian. For example, mesoscopic fluctuations in the orbital magnetization of a chaotic grain in the presence of an orbital magnetic field were studied in Ref.~\onlinecite{murthy}. Our studies here are focused on grains with equally-spaced spectra and do not include the effects of mesoscopic fluctuations.

The outline of this paper is as follows. In Sec.~\ref{afmc} we discuss the AFMC method and its particular realization for the BCS pairing Hamiltonian. The method is based on a Hubbard-Stratonovich transformation,~\cite{hstrans} and the number of particles is fixed using a particle-number projection. Section \ref{renorm} discusses various ways to renormalize the pairing coupling constant when the band width is reduced. By reducing the band width, the low-temperature AFMC calculations can be carried out more efficiently. In Sec.~\ref{results}, we use AFMC to calculate the spin susceptibility and heat capacity of grains with different values of $\Delta/\delta$.
\section{Auxiliary-field Monte Carlo method}\label{afmc}

We briefly discuss the model and then describe the application of the AFMC method to an ultra-small metallic grain.

\subsection{Model}
We describe the metallic grain by the BCS pairing
Hamiltonian \cite{tichy}
\begin{eqnarray}
\label{origH}
H=\hspace{-0.2cm}\sum_{i=-N_o}^{N_o}\hspace{-0.2cm}\epsilon_i\left(c_{i+}^\dagger
c_{i+}+c_{i-}^\dagger
c_{i-}\right)-g\hspace{-0.2cm}\sum_{i,j=-N_o}^{N_o}\hspace{-0.2cm}c_{i +}^\dagger
c_{i-}^\dagger c_{j-}c_{j+}\,,
\end{eqnarray}
where $c_{i\sigma}^\dagger$ is the creation operator for an
electron in the single-particle level $i$ with either spin up
($\sigma=+$) or spin down ($\sigma=-$). We assume $2 N_o +1$ levels with a cutoff $N_o$
determined by the Debye frequency $\omega_D$ via $N_o \approx \omega_D/\delta$ where $\delta$  the mean single-particle level spacing. The second term in
(\ref{origH}) is a pairing interaction, which scatters only
time-reversed pairs of electrons. In the following we will
assume an equidistant spectrum $\epsilon_i=i\delta$ ($i=-N_o,\ldots,N_o$) and half filling, i.e., the number of electrons is either $N=2 N_o$ (even) or $N=2N_o+1$ (odd). The coupling constant $g$ is related to
the BCS pairing gap (at zero temperature) through
$\Delta=2\omega_D e^{-\delta/g}$. Both the single-particle mean level spacing and the pairing gap are assumed to be small compared with the Debye frequency, $\delta,\Delta \ll \omega_D$.

\subsection{Hubbard-Stratonovich transformation}
The BCS Hamiltonian (\ref{origH}) can be written in a
density decomposition as
\begin{eqnarray}\label{density-H}
H=\sum_i\epsilon_i \hat n_i-{g\over 4}\sum_{ij}
[(\rho_{ij}+\bar{\rho}_{ij})^2\ -(\rho_{ij}-\bar{\rho}_{ij})^2] \;,
\end{eqnarray}
where $\hat n_i=c_{i+}^\dagger c_{i+}+c_{i-}^\dagger c_{i-}$ is the occupation number operator of level $i$, $\rho_{ij}=c_{i+}^\dagger c_{j+}$ are spin-up density operators and $\bar{\rho}_{ij}=c_{i-}^\dagger c_{j-}$  is the time reverse operator of $\rho_{ij}$ describing a spin-down density operator.

The operator $e^{-\beta H}$ at inverse temperature $\beta=1/T$  can be viewed as a many-body propagator in imaginary time $\beta$. In the Hubbard-Stratonovich (HS) transformation it can be written as a path integral over one-body propagators that describe non-interacting electrons moving in external time-dependent auxiliary fields.
This is accomplished by dividing the time interval $(0,\beta)$ into $N_t$ time slices of length $\Delta\beta$ each such that $e^{-\beta H}=\left(e^{-\Delta\beta H}\right)^{N_t}$. For each time slice $e^{-\Delta\beta H} \approx e^{-\Delta \beta \sum_i \epsilon_i \hat n_i} \Pi_{ij} e^{-\Delta \beta {g\over 4} [(\rho_{ij}+\bar{\rho}_{ij})^2\ -(\rho_{ij}-\bar{\rho}_{ij})^2]}$ to order $(\Delta\beta)^2$. The interaction terms in the exponent appear as sum of squares and the propagator for one time slice can be written as Gaussian integrals over real auxiliary fields $\sigma^R_{ij}$ and $\sigma^I_{ij}$
\begin{eqnarray}
&& e^{-\Delta \beta {g\over 4} \left[(\rho_{ij}+\bar{\rho}_{ij})^2\ -(\rho_{ij}-\bar{\rho}_{ij})^2\right]} =
{\beta g \over 4 \pi} \int d\sigma^R_{ij} d\sigma^I_{ij} \\ \nonumber  && \;\;\;\; \times e^{-\Delta\beta\frac{g}{4}  \left[(\sigma^R_{ij})^2  + (\sigma^I_{ij})^2\right]} e^{-\Delta\beta {g\over 2} \sigma^R_{ij} \left[(\rho_{ij}+\bar{\rho}_{ij}) + i \sigma^I_{ij} (\rho_{ij}-\bar{\rho}_{ij})\right]} \;.
\end{eqnarray}

 Using different complex fields at each time slice
$\sigma_{ij}(\tau_n)= \sigma^R_{ij}(\tau_n) + i\sigma^I_{ij}(\tau_n)$ ($\tau_n=n\Delta \beta$ with $n=1,\ldots,N_t$),
the propagator $e^{-\beta H}$ can be written as a path integral
\begin{eqnarray}\label{hs-transformation}
 e^{-\beta H}=\int {\cal
D}[\sigma]G(\sigma)U_\sigma(\beta,0)
\end{eqnarray}
with the measure
\begin{eqnarray}
{\cal
D}[\sigma]=\prod_{ij,n}\left[{d\sigma_{ij}(\tau_n)d\sigma_{ij}^*(\tau_n)\over 2i} {\Delta\beta
g \over 4\pi} \,\right]\;,
\end{eqnarray}
and  Gaussian factor
\begin{eqnarray}
G(\sigma)=\exp\left[-{g\over 4}\,\Delta\beta \sum_{ij,n}
|\sigma_{ij}\left(\tau_n\right)|^2\right]\;.
\end{eqnarray}
$U_\sigma$ in (\ref{hs-transformation}) describes the propagator of non-interacting electrons in external time-dependent auxiliary fields $\sigma_{ij}\left(\tau\right)$
\begin{eqnarray}
\label{prop}
U_\sigma\left(\beta,0\right)=\,e^{-\Delta\beta h_\sigma\left(\tau_{N_t}\right)}\ldots e^{-\Delta\beta h_\sigma\left(\tau_1\right)}
\end{eqnarray}
with a time-dependent one-body Hamiltonian
\begin{eqnarray}\label{h-sigma}
h_\sigma(\tau) =\sum_i\epsilon_i \hat n_i
-\frac{g}{2}\sum_{ij}\left[\rho_{ij}\,\sigma_{ij}^*(\tau)
+\bar{\rho}_{ij}\,\sigma_{ij}(\tau)\right] \;.
\end{eqnarray}

Eq.~(\ref{hs-transformation}) is known as a Hubbard-Stratonovich (HS) transformation.

 Another HS transformation can be obtained using good-spin density operators
\begin{eqnarray}
\rho_{K M}\left(ij\right)=\sum_{\sigma_i\sigma_j}\left({1/2},{\sigma_i/2};{1/2},{\sigma_j/2}|K
M\right)c_{i\sigma_i}^\dagger \tilde{c}_{j\sigma_j}
\end{eqnarray}
with $\tilde{c}_{i\sigma_i}=(-1)^{(1+\sigma_i)/2}c_{i-\sigma_i}$ and
$K= 0,1$; $-K \leq M \leq K$. We rewrite the pairing Hamiltonian (\ref{origH})
in terms of these good-spin density operators as
\begin{eqnarray}\label{good-spin}
H =\sum_i\left(\epsilon_i-{g\over 4}\right) \hat
n_i - {g\over 4}\sum_{ij \atop K M}\rho_{K
M}\left(ij\right)\bar\rho_{K M}\left(ij\right)\;,
\end{eqnarray}
where $\bar \rho_{K M} = (-1)^{K+M} \rho_{K -M}$ is the time-reversed operator of $\rho_{K M}$. Eq.~(\ref{good-spin}) can be rewritten in a form similar to (\ref{density-H}) but now with $\rho_{KM}(ij)$ replacing $\rho_{ij}$. An HS decomposition can be carried out, introducing the complex fields $\sigma_{KM}(ij)$.  This HS decomposition is similar to the one used in SMMC.

To account for finite-size effects in the grain, it is necessary to evaluate thermal expectation values in the canonical ensemble with fixed electron number $N$.  For an observable $O$
\begin{eqnarray}
\label{eq:obs} \langle O\rangle\equiv {\Tr_N(O e^{-\beta H})\over \Tr_N e^{-\beta H}}={\int
{\cal D}[\sigma]W_\sigma \Phi_\sigma\langle O\rangle_\sigma\over
\int {\cal D}[\sigma]W_\sigma \Phi_\sigma}\,,
\end{eqnarray}
where $\Tr_N$ denotes a trace at a fixed number of particles $N$. Here
$W_\sigma=G(\sigma)|\Tr_N U_\sigma(\beta,0)|$ is a positive definite weight function,
$\Phi_\sigma=\Tr_N U_\sigma(\beta,0)/|\Tr_N U_\sigma(\beta,0)|$ is the ``sign'' function, and
 $\langle O\rangle_\sigma=\Tr_N [
OU_\sigma(\beta,0)]/\Tr_N U_\sigma(\beta,0)$. Both $\Tr_N U_\sigma$ and $\langle O\rangle_\sigma$ can be evaluated using
matrix algebra in the single-particle space. For example, the
grand-canonical trace of the one-body propagator $U_\sigma(\beta,0)$ in Fock space is
given by
\be\label{Tr-U-sigma}
\Tr U_\sigma(\beta,0)=\det [1+{\bf U}_\sigma(\beta,0)] \;,
\ee
where ${\bf U}_\sigma(\beta,0)$ is the $(4N_o+2)\times (4N_o+2)$ matrix
representing $U_\sigma(\beta,0)$ in the single-particle space (each level is spin degenerate and thus the dimension of the single-particle space is twice the number of levels).
\subsection{Particle-number projection}
The canonical trace can be evaluated from $\Tr_N \hat X=\Tr (P_N \hat X)$ where
$P_{N}=\delta(\hat{N}-N)$ is the particle-number projector. Since the number operator
$\hat{N}$ takes a finite number of discrete values $0,1,\cdots,4N_o+2$, $P_N$ can be
written as a discrete Fourier transform
\be
P_{N}={1\over 4N_o+2}\sum_{m=1}^{4N_o+2}
e^{-i\phi_m N} e^{i\phi_m \hat{N}}  \;,
\ee
 where $\phi_m=2\pi m/(4N_o+2)$ are quadrature points. The canonical trace of $U_\sigma$ is
then given by~\cite{ormand,alhassid00}
\begin{eqnarray}
 \Tr_N U_\sigma(\beta,0) & = & {1\over
4N_o+2}\sum_{m=1}^{4N_o+2} e^{-i\phi_m N} \\\nonumber
&&\;\;\; \times
\det[1+e^{i\phi_m}{\bf U}_\sigma(\beta,0)]\;.
\end{eqnarray}
Similarly for a one-body observable
\begin{eqnarray}
\label{cantrace} && \Tr_N [ OU_\sigma(\beta,0)] ={1\over
4N_o+2}\sum_{m=1}^{4N_o+2} e^{-i\phi_m N} \\\nonumber
&&\times \det[1+e^{i\phi_m}{\bf U}_\sigma(\beta,0)]
{\rm tr}\left({1\over {\bf 1} +e^{-i\phi_m}{\bf
U}_\sigma^{-1}(\beta,0)}{\bf O}\right) \;,
\end{eqnarray}
where ${\bf O}$ is the matrix representing the operator $O$ in the single-particle space.
\subsection{Monte Carlo sign}
 The functional integral in (\ref{eq:obs}) is a multi-dimensional
integral over a large number of auxiliary fields. In the AFMC method, this integration is carried out by Monte Carlo methods. The auxiliary fields are sampled according to the
distribution $W_\sigma$, and for $M$ samples $\{\sigma_i\}$, the thermal expectation value (\ref{eq:obs}) is estimated by
\begin{eqnarray}
\label{eq:obsmc}\langle O\rangle\approx {{1\over M}\sum_i
\Phi_{\sigma_i}\langle O\rangle_{\sigma_i} \over {1\over M}\sum_i
\Phi_{\sigma_i}} \;.
\end{eqnarray}

In general, the sign function $\Phi_\sigma$ is not necessarily positive and fluctuates from sample to sample. When the statistical error of ${\rm Re}\Phi_\sigma$ is larger than its average value, the observables cannot be reliably estimated, a problem known as the Monte Carlo sign problem. In the case of the attractive pairing interaction, the single-particle Hamiltonian (\ref{h-sigma}) is time-reversal invariant, $\bar h_\sigma = h_\sigma$, and so is $U_\sigma$. Therefore the eigenvalues of
${\bf U}_\sigma(\beta,0)$ come in complex conjugate pairs (corresponding to an eigenstate and its time-reversed state). Thus the grand-canonical many-particle partition $\Tr
U_\sigma(\beta,0)$ in (\ref{Tr-U-sigma}) is positive definite. When projected on an even number of electrons $\Tr_N
U_\sigma(\beta,0)$ remains almost always positive and there is no sign problem. When projected on an odd number of electrons $\Tr_N
U_\sigma(\beta,0)$ can be negative and a sign problem occurs at large values of $\beta$. In practice, we use the reprojection method.~\cite{ALN99} We carry out the Monte Carlo sampling with an even number of electrons and then reproject on an odd number of electrons.
\subsection{Discretization effects}

The discretization of imaginary time introduces a systematic error in the
evaluation of the HS integral (\ref{hs-transformation}).
We corrected for these discretization effects by carrying out AFMC calculations for several values of $\Delta \beta$ (usually  $1/32$
and $1/64$ $\delta^{-1}$) and then performing a linear extrapolation to $\Delta\beta=0$.

\section{Renormalization}\label{renorm}
The AFMC method has the advantage that it only requires matrix algebra in
single-particle space. The computational time scales
as a power law $\sim N_o^4$ with the number of single-particle
states and not exponentially as is the case with conventional diagonalization methods. We
can further reduce the computational effort at low temperatures by
using the Hamiltonian (\ref{origH}) in a truncated model space with
a smaller cutoff $N_r < N_o$ and a renormalized coupling constant
$g_r$.  For a renormalization group (RG) approach that describes the BCS instability in the bulk, see Ref.~\onlinecite{shankar}. In this section we discuss various ways to determine $g_r$ for a finite-size system.

In the bulk, thermodynamic quantities are expected to be universal functions of $T/\Delta$ where $\Delta$ is the BCS gap. The $T=0$ BCS gap is determined from $1/g=\sum_k 1/(2 E_k)$ where $E_k=\sqrt{(\epsilon_k-\mu)^2+\Delta^2}$ are the quasi-particle energies. Taking the bulk limit $\delta \to 0$ of a picketfence spectrum in a band between $-\omega_D$ and $\omega_D$ with half filling ($\mu \approx 0$), the solution for $\Delta$ is given by $\Delta = \omega_D/\sinh(1/\lambda)$ where $\lambda=g/\delta$, or  $\Delta = 2\omega_D e^{-1/\lambda}$ for $\lambda \ll 1$.
An effective Hamiltonian in a truncated flat band of half width $D$ (smaller than $\omega_D$) can be obtained by renormalizing the coupling constant $\lambda$ to a value $\lambda_r$ such that the gap parameter remains fixed, i.e., $\Delta = D/\sinh(1/\lambda_r)$.

In a finite-size system, the discreteness of the single-particle spectrum becomes important and there are two energy scales: the BCS gap $\Delta$ and the single-particle mean level spacing $\delta$. When truncating to a band with $2N_r+1$ levels, we renormalize $g$ such that (at fixed $\delta$) the BCS gap $\Delta$, determined by the BCS equation for a discrete spectrum, remains fixed. Using the gap equation for both untruncated and truncated spectra, we find the following relation between the bare coupling constant $g$ and the renormalized constant $g_r$
\be\label{renorm1}
{1\over g} \approx {1\over g_r} + \sum_{N_r <|k|\leq N_o} {1\over 2 |\epsilon_k -\mu|} \;.
\ee
In the sum within the shell $N_r <|k|\leq N_o$ (outside the truncated band), we have ignored the gap $\Delta$ assuming $\Delta \ll N_r\delta$. We have also assumed that the chemical potential is the same for $N_o$ levels and for $N_r$ levels. For the picketfence spectrum at half filling this is exact by symmetry; $\mu=-\delta/2$ for an even number of particles $N$ and $\mu=0$ for odd $N$.

Equation (\ref{renorm1}) can be rewritten as
\be\label{halperin-ren}
g_r = g \left( 1 - g \sum_{N_r <|k|\leq N_o} {1\over 2 |\epsilon_k -\mu|}\right)^{-1} \;.
\ee
This equation was also obtained using a perturbative treatment~\cite{halperin} by taking the limit $g\rightarrow 0$ and
$\omega_D\rightarrow\infty$ at fixed $\Delta$ and $\delta$. A quite accurate estimate of the discrete sum in (\ref{halperin-ren}) for a picketfence spectrum is given by (for $\mu=0$) $\sum_{N_r < k\leq N_o} 1/|k| \approx \int_{N_r+1/2}^{N_o +1/2}dx/x = \ln[(2N_o+1)/(2N_r+1)]$. Using this relation in (\ref{halperin-ren}) and $\Delta =2\omega_D e^{-\delta/g}$ with $\omega_D=(N_o+1/2)\delta$, we arrive at
\be\label{bulk-ren}
{g_r \over \delta} = {1 \over \ln\left({2N_r+1 \over \Delta/\delta}\right)} \;.
\ee

\begin{figure}[t]
\epsfxsize=0.90\columnwidth \epsfbox{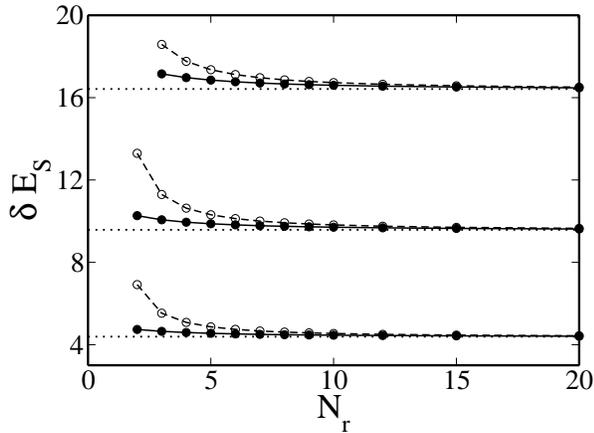}
\caption{\label{spin_gaps} The excitation gaps $\delta E_S$ (i.e., lowest excitation energy for a given spin $S$) for spins $S=1/2,/3/2,5/2$ versus the half width $N_r$ of the truncated band. The gaps are shown for $\Delta/\delta=3$. Solid circles correspond to keeping the discrete BCS gap fixed while open circles correspond to the renormalization~\cite{halperin} of Eq.~(\ref{halperin-ren}). Dotted lines describe the asymptotic values of $\delta E_S$ for large $N_r$.}
\end{figure}

For a picketfence spectrum, rather than using (\ref{halperin-ren}), we can calculate quite accurately the discrete sum in the BCS gap equation using $1/g =\sum_{k \leq |N_r|} 1/(2 E_k) \approx \int_{-N_r-1/2}^{N_r+1/2} dx /\sqrt{x^2+ (\Delta/\delta)^2}$. We find
\be\label{sinh-ren}
{g_r \over \delta} = {1 \over {\rm arcsinh}\left({N_r+1/2 \over \Delta/\delta}\right)} \;.
\ee
\begin{figure}[t]
\epsfxsize=0.95\columnwidth \epsfbox{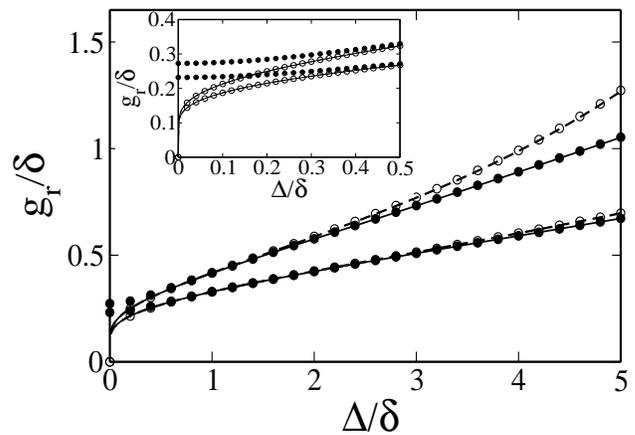}
\caption{\label{renormalization} Renormalized coupling constant $g_r$ versus $\Delta/\delta$ for $N_r=5$ (upper curves) and $N_r=10$ (lower curves). Solid circles correspond to keeping the discrete BCS gap $\Delta$ fixed  (at fixed $\delta$) and open circles describe the renormalization of Eq.~(\ref{halperin-ren}). Solid lines correspond to (\ref{sinh-ren}) and dashed lines are the bulk renormalization (\ref{bulk-ren}).  The inset shows the region $\Delta/\delta < 0.5$.}
\end{figure}
Figure \ref{renormalization} compares the various renormalized values of the coupling constant $g_r$  versus $\Delta/\delta$ for truncated bands of $N_r=5$ (upper curves) and $N_r=10$ (lower curves). Open circles describe the renormalization of Eq.~(\ref{halperin-ren}), while solid circles correspond to keeping the discrete BCS gap fixed. Dashed lines are the bulk renormalization (\ref{bulk-ren}) and solid lines describe the normalization given by (\ref{sinh-ren}). We observe that for $\Delta/\delta > 0.5$, the renormalization (\ref{sinh-ren}) essentially coincides with keeping the discrete BCS gap fixed, while the bulk renormalization (\ref{bulk-ren}) coincides with the renormalization (\ref{halperin-ren}) of Ref.~\onlinecite{halperin}. The renormalization (\ref{halperin-ren}) deviates from the discrete BCS renormalization at large values of $\Delta/\delta$ but this deviation becomes smaller as $N_r$ increases. This is consistent with the condition $\Delta \ll N_r \delta$. A different behavior is observed for $\Delta/\delta \leq 0.5$ (see inset of Fig.~\ref{renormalization}) where the renormalization (\ref{sinh-ren}) coincides with (\ref{halperin-ren}).
To test the different renormalization methods, we show in Fig.~\ref{spin_gaps} the excitation gaps $\delta E_S$ for spins $S=1/2, 3/2$ and $5/2$ (for an odd number of particles) as a function of the width $N_r$ of the truncated band for a fixed value of $\Delta/\delta=3$. These spin gaps were determined by solving Richardson's equations.  Open circles correspond to Eq.~(\ref{halperin-ren}) of Ref.~\onlinecite{halperin} while solid circles correspond to keeping the discrete BCS gap fixed. We observe faster convergence for the second method. However, the discrete BCS equation is meaningful as long as $\Delta/\delta \geq 0.5$. For $\Delta/\delta < 0.5$,  the method of Ref.~\onlinecite{halperin} (or alternatively Eq.~(\ref{sinh-ren}) for a picketfence spectrum) should be used and it gives the correct limit $g_r\to 0$ when $\Delta/\delta \to 0$.

Figure \ref{spin_gaps} also demonstrates that the $T=0$ renormalization of the coupling constant approximately preserves the low-energy excitation spectrum of the grain. Thus, the same renormalization is expected to work well at finite temperature $T$, as long as $N_r$ is sufficiently large. At higher temperatures, higher-energy configurations become populated and it is necessary to increase the band width $N_r$. In our calculations we ensure
that our model space is large enough by estimating the minimal
required value of $N_r$ at each temperature from the canonical Fermi gas
results. We also note that the renormalization procedure works well without introducing new coupling constants beyond $g$.  For grains with equally-spaced spectra and given $\Delta/\delta$, the various thermodynamic properties are expected to be universal functions of $T/\delta$, irrespective of the bandwidth.

As an example we consider an ultra-small aluminum grain with
$\Delta/\delta=1$, where $\Delta$ denotes the bulk pairing
gap at zero temperature. The measured Debye frequency is
$\omega_D\approx 34 \,{\rm meV}$. Using the experimental value for
the zero-temperature gap of thin films $\Delta\approx 0.38\,{\rm
meV}$, we can estimate the bare cutoff to be $N_o\approx 89$. The bare coupling constant is then given by $g/\delta \approx 1/\ln(2\omega_D/\Delta) = 0.193$. In the actual calculations, we used truncated values $N_r=10,15,25$ (depending on temperature) and used the appropriate  renormalized values $g_r$ of the coupling constant.

\section{Thermodynamic properties}\label{results}

We used AFMC to calculate thermodynamic properties of a metallic grain in the crossover regime. In particular, we calculated the spin susceptibility and heat capacity for grains with both even and odd number of electrons.

\subsection{Spin Susceptibility}
The spin susceptibility is defined as the magnetic response of the grain to an external Zeeman field $H$. Taking the $z$ axis along the field direction, the field couples directly to the spin projection $S_z$, and the spin susceptibility is given by
\begin{eqnarray}
\label{def1} \chi(T)= \left.{\partial M \over \partial H}\right|_{H=0}= 4 \beta \mu_{\rm B}^2\left(\langle S_z^2\rangle - \langle S_z\rangle^2\right) \;,
\end{eqnarray}
where $M$ is the magnetization of the grain, $\mu_B$ is the Bohr magneton and we have used a  $g$-factor of 2 for free electrons.  The spin projection is given by
$S_z=\sum_{i}(c_{i+}^\dagger
c_{i+}-c_{i-}^\dagger c_{i-})/2$.

\begin{figure}[t]
\epsfxsize=0.95\columnwidth\epsfbox{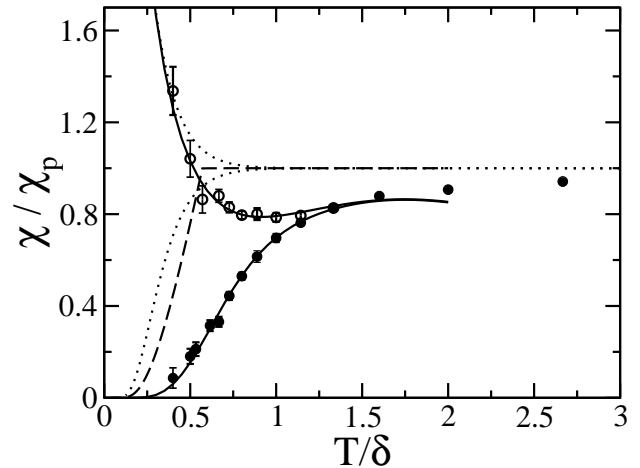}
\caption{\label{sus_1}Spin susceptibility $\chi$ (in units of the Pauli susceptibility $\chi_p=2 \mu_B^2/\delta$) versus $T/\delta$ for $\Delta/\delta=1$.
Solid circles are the AFMC results for an even number of electrons and open circles are AFMC results for an odd number of electrons. The solid lines are obtained by using Richardson solution for all energy levels below an excitation energy of $\sim 30 \delta$. For comparison we also show the spin susceptibility of the free Fermi gas in the canonical ensemble (dottes
lines), and of the BCS mean-field theory (dashed line).}
\end{figure}

AFMC results for the spin susceptibility of a grain with  $\Delta/\delta=1$ are shown in Fig.~\ref{sus_1}. Solid circles (with statistical errors) are for even number of electrons and open circles are for odd number of electrons (at half filling). The spin susceptibility is measured in units of the Pauli susceptibility  $\chi_p=2\mu_{\rm B}^2/\delta$ (see below).  We used $N_r=10$ and $M=16000$ samples
for $T\le 0.8\,\delta$ (at $T=0.4\,\delta$ we have used $M=64000$
samples), $N_r=15$ and $M=4000$ samples for $0.8\,\delta \leq T\le 1.5\,\delta$,
$N_r=25$ and $M=4000$ at higher temperatures. The larger number of samples at lower temperatures is necessary to overcome a mild sign problem for the reprojection on the odd number of electrons. We also solved Richardson's equations using the method of Ref.~\onlinecite{rombouts} up to an
excitation energy of $30\,\delta$, and the results are shown by the solid lines in Fig.~\ref{sus_1}. They agree with the AFMC results up to
 $T \approx 1.25\,\delta$. At higher temperatures, it is necessary to increase
the  maximal excitation energy. However, this becomes a difficult problem as the number of many-body levels proliferates combinatorially with excitation energy.~\cite{richardson}

\begin{figure}[t]
\epsfxsize=0.95\columnwidth\epsfbox{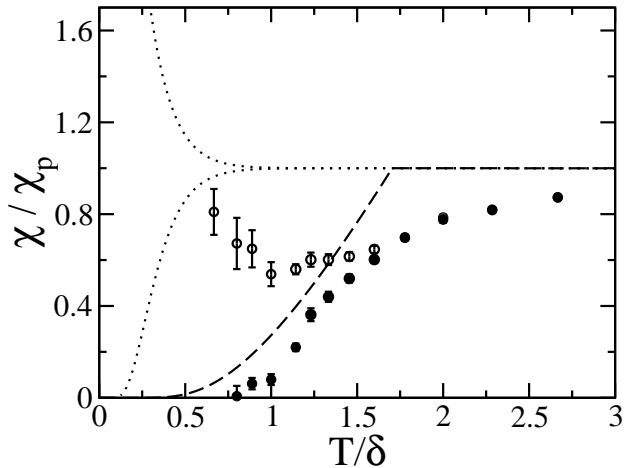}
\caption{\label{sus_3} Spin susceptibility versus $T/\delta$ for $\Delta/\delta=3$.
Notation is as in Fig.~\ref{sus_1}.}
\end{figure}

For comparison, we have also calculated the spin susceptibility of a free Fermi gas in the canonical ensemble. The results are shown by the dotted lines in Fig.~\ref{sus_1} for even and odd number of electrons.

At high temperatures, we can calculate $\chi$ for a free Fermi gas in the grand-canonical ensemble. We have
\begin{eqnarray}\label{chi-fermi}
\label{chi-c} \chi(T)=2 \beta \mu_{\rm B}^2\sum_i
f_i (1- f_i)\,
\end{eqnarray}
where $f_i = 1/\left[1+e^{\beta(\epsilon_i-\mu)}\right]$ is the Fermi-Dirac occupation of level $i$. For the half-filled picket-fence spectrum, $\mu=0$, and for temperatures $T \gg \delta$ (but much smaller than the Fermi energy)
\begin{equation}\label{Pauli}
\chi \approx {2\mu_B^2 \over \delta} \int_{-\infty}^{\infty} dx {e^x \over (1+e^x)^2} = {2\mu_B^2 \over \delta} \;.
\end{equation}
This value of $\chi$ is the Pauli susceptibility $\chi_p$. We observe that the canonical Fermi gas susceptibility (dotted lines in Fig.~\ref{sus_1}) essentially coincides with (\ref{Pauli}) (i.e., $\chi \approx \chi_p$) already for $T/\delta >1$. The AFMC results (that includes the pairing interaction) approaches Pauli's limit at high temperatures.

In the BCS limit, we can calculate $\chi$ from (\ref{chi-fermi}) but now $f_i$ are the quasi-particle occupation numbers
$f_i=1/\left(1+e^{\beta E_i}\right)$ with $E_i=\sqrt{(\epsilon_i-\mu)^2 +\Delta^2}$ being the quasi-particle energies. The BCS spin susceptibility is shown by the dashed line in Fig.~\ref{sus_1}. We observe that pairing correlations in $\chi$ persist up to temperatures that are higher than the BCS critical temperature.

For larger gap values, even-odd effects extend to higher temperatures and it is difficult to use Richardson method. On the other hand, AFMC calculations remain tractable. The AFMC spin susceptibility for  $\Delta/\delta=3$ is shown in Fig.~\ref{sus_3}. Here we used
$N_r=10$ and $M=64000$ samples for $T\le 1.25\,\delta$ (at
$T=0.67\,\delta$ we have used $M=384000$ samples), $N_r=15$ and
$M=16000$ samples for $ 1.25\,\delta \leq T\le 1.6\,\delta$, $N_r=25$ and $M=4000$ at higher temperatures.

\begin{figure}[t!]
\epsfxsize=0.95\columnwidth\epsfbox{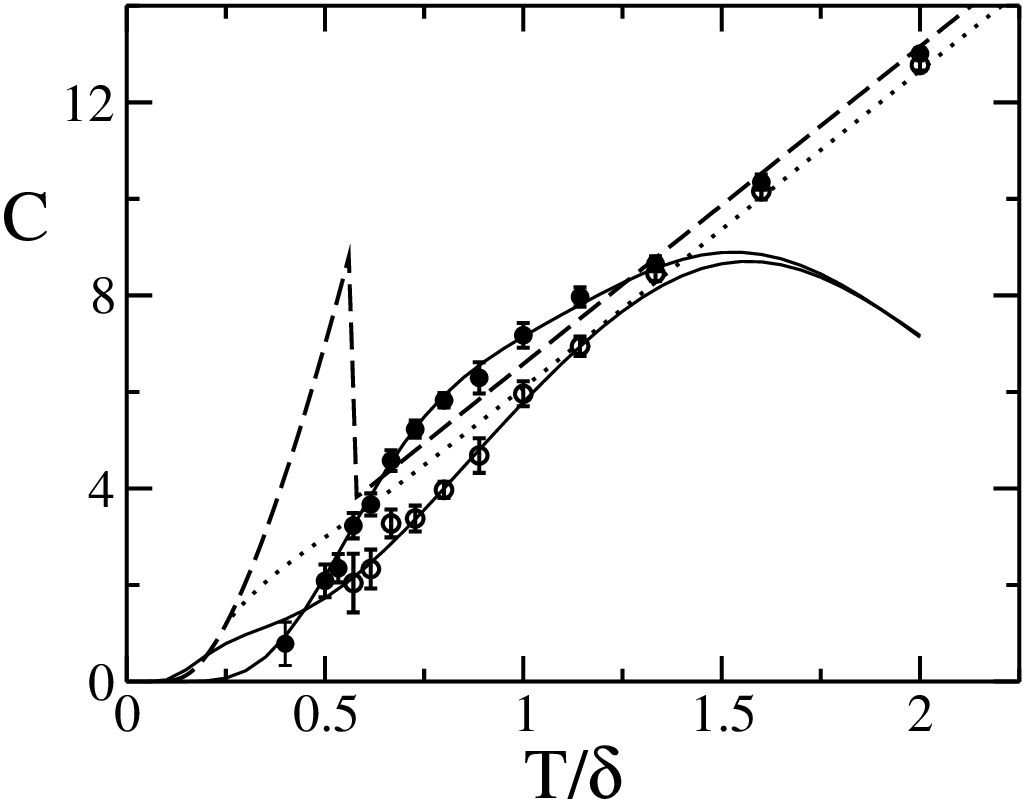}
\caption{\label{hc_1}Heat capacity for $\Delta/\delta=1$. See
Fig.~\ref{sus_1} for notation.}
\end{figure}
We observe the following:
\begin{itemize}

\item The pairing interaction suppresses the
spin susceptibility when compared with the canonical free fermi gas susceptibility for both even and odd number of electrons. At
higher temperatures, the results for the even and odd number of particles coincide and
approach asymptotically the Pauli spin susceptibility.

\item The spin susceptibility for an odd number of electrons shows a
characteristic reentrant behavior, in agreement with findings in Refs.~\onlinecite{lorenzo} and \onlinecite{rombouts}. The unpaired electron leads to a Curie-like divergence $\chi\sim 1/T$ in the limit $T\rightarrow 0$ (even for the canonical Fermi gas). However,  pairing correlations suppress the spin susceptibility, leading to a minimum in $\chi$ at a certain temperature. This behavior of $\chi$ is similar to the the behavior observed in the moment of inertia of an odd-even or odd-odd nucleus.~\cite{inertia}

\item In the presence of pairing correlations, number-parity (i.e., even-odd) effects already appear at temperatures $T\le 1.5\,\delta$ for $\Delta/\delta=1$, as compared to
$T\le 1.0\,\delta$ for the free Fermi gas. For larger values of $\Delta/\delta$, the region of reentrant behavior is shifted to higher temperatures and is more pronounced. The spin susceptibility for an even number of electrons gets closer to its BCS approximation at larger values of $\Delta/\delta$.

\end{itemize}

\subsection{Heat Capacity}

\begin{figure}[t!]
\epsfxsize=0.95\columnwidth\epsfbox{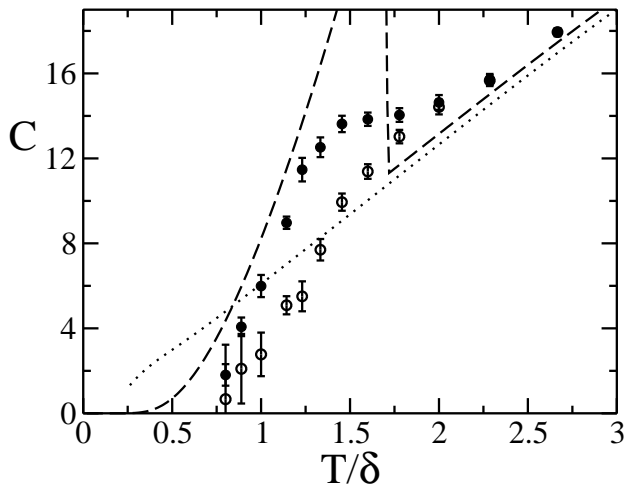}
\caption{\label{hc_3} Heat capacity for $\Delta/\delta=3$. See
Fig.~\ref{sus_1} for notation.}
\end{figure}

The heat capacity $C= dE/dT $ is calculated in AFMC as a numerical derivative
of the thermal energy $E(\beta)=\Tr (H e^{-\beta H})/ \Tr e^{-\beta H}$. We use
\begin{eqnarray}
\label{deriv}
  C = -\beta^2\frac{E(\beta +\delta \beta ) -
    E(\beta -\delta \beta )}{2\delta \beta } + O(\delta \beta )^{2} \;.
\end{eqnarray}
The statistical errors are reduced by using the same set of
auxiliary fields $\sigma$ for the calculation of both
 $E(\beta \pm\delta \beta )$ and taking into account correlated errors.~\cite{liu}
Figures \ref{hc_1} and \ref{hc_3} show the AFMC heat capacity for $\Delta/\delta=1$ and $\Delta/\delta=3$, respectively, for both even (solid circles) and odd (open circles) number of electrons. The solid lines in Fig.~\ref{hc_1} are Richardson results when energy levels up to $\sim 30 \,\delta$ are included. They agreed with AFMC for $T \leq 1.3\,\delta$ but deviate at higher temperatures because of truncation effects.

For a non-interacting
Fermi gas in the grand-canonical ensemble, we calculate the heat capacity from $C=TdS/dT$, where the entropy $S$ is given by
\begin{eqnarray}\label{entropy}
 S=-2\sum_{i}\left[ f_i\ln
f_i + (1- f_i)\ln(1- f_i)\right] \;.
\end{eqnarray}
The factor of $2$ in (\ref{entropy}) accounts for spin degeneracy.
 In the high-temperature limit $T\gg\delta$ and for a picketfence spectrum, we find
 \begin{equation}\label{hc_fermi}
  C \approx {2 \over \beta \delta} \int_{-\infty}^{\infty} dx {x^2 e^x \over (1+e^x)^2} = {2\pi^2 \over 3}{T\over \delta} \;.
\end{equation}
 The canonical Fermi gas heat capacity (for an even number of electrons) is shown by the dotted lines in Figs.~\ref{hc_1} and \ref{hc_3}. The BCS heat capacity is calculated from $C=TdS/dT$ using the quasi-particle occupations for $f_i$ in (\ref{entropy}) and is shown by the dashed lines.

We observe the following:
\begin{itemize}
\item For intermediate temperatures, the heat capacity is enhanced for an even
number of particles as compared with the heat capacity for an odd number
of particles. This effect is clearly seen for $\Delta/\delta=3$ where the heat capacity for an even number of electrons displays a shoulder. A similar behavior was found in the heat capacity of even-even neutron-rich nuclei.~\cite{liu,level} This even-odd effect is a signature of pairing correlations in the crossover regime.  At very low temperatures this behavior is reversed and the heat capacity
is more strongly suppressed in the even case (see the solid lines in Fig.~\ref{hc_1}).

\item The AFMC heat capacity is suppressed with respect to the BCS heat capacity. For $\Delta/\delta=3$, the shoulder in the heat capacity for the even number of electrons occurs around the BCS critical temperature. However, for $\Delta/\delta=1$, this shoulder structure occurs at higher temperatures than the BCS critical temperature. This suggests that the even-odd effects occur on the scale that is the larger between $\Delta$ and $\delta$.
\end{itemize}

\section{Conclusion}

We used an auxiliary-field Monte Carlo approach to calculate thermal observables of ultra-small metallic grains. The method is particularly useful for grains in the crossover regime between the BCS limit and the fluctuation-dominated limit, where the BCS gap is comparable to the single-particle mean-level spacing. For an attractive pairing interaction, there is no Monte Carlo sign problem for an even number of electrons, and accurate calculations are possible. The computational effort can be further reduced by renormalizing the interaction in a truncated band. For a picketfence spectrum and for a given ratio $\Delta/\delta$ of the pairing gap to the mean level spacing, the spin susceptibility and heat capacity are universal functions of $T/\delta$.

We thank Y. Imry, R. Shankar and J. von Delft for useful discussions. This work was supported in part by the U.S. DOE grant No.\ DE-FG-0291-ER-40608.

\vfill

\begin{thebibliography}{99}
%
\bibitem{BCS} J. Bardeen, L.N. Cooper and J.R. Schrieffer,
Phys. Rev. {\bf 108}, 1175 (1957).
%
\bibitem{vonDelft}
J.\ von Delft, Ann.\ Phys. (Leipzig) {\bf 10}, 2001; J.\ von Delft
and D.\ C.\ Ralph, Phys. Rep. {\bf 345}, 661-173 (2001) and references therein.
%
\bibitem{RBT}
D.\ C.\ Ralph, C.\ T.\ Black, and M.\ Tinkham, Phys.\ Rev.\ Lett
{\bf 74}, 3241 (1995); C.\ T.\ Black, D.\ C.\ Ralph, and M.\
Tinkham, Phys.\ Rev.\ Lett {\bf 76}, 688 (1996); D.\ C.\ Ralph,
C.\ T.\ Black, and M.\ Tinkham, Phys.\ Rev.\ Lett {\bf 78}, 4087
(1997).
%
\bibitem{tichy}
J.\ von Delft, A.\ D.\ Zaikin, D.\ S.\ Golubev, and M.\ Tichy,
Phys.\ Rev.\ Lett.\ {\bf 77}, 3189 (1996)
%
\bibitem{lorenzo}
A.\ Di\ Lorenzo, R.\ Fazio, F.\ W.\ J.\ Hekking, G.\ Falci, A.\
Mastellone, and G.\ Giaquinta, Phys.\ Rev.\ Lett {\bf 84}, 550
(2000).
%
\bibitem{richardson}
R.\ W.\ Richardson, Phys.\ Rev.\ Lett. {\bf 3}, 277 (1963); R.\
W.\ Richardson, Phys.\ Rev.\ {\bf 159}, 792 (1967).
%
\bibitem{SPA} B. Muhlschlegel, D.J. Scalapino and R. Denton,
Phys. Rev. B {\bf 6}, 1767 (1972); Y. Alhassid and J. Zingman,
Phys. Rev.  {\bf C30}, 684 (1984); B. Lauritzen, P. Arve and G.F. Bertsch, Phys.
Rev. Lett. {\bf 61}, 2835 (1988).
%
\bibitem{schechter} M. Schechter, Y. Imry, Y. Levinson and J. von Delft, Phys. Rev. B {\bf 63}, 214518 (2001).

\bibitem{smmc}
G.\ H.\ Lang, C.\ W.\ Johnson, S.\ E.\ Koonin, and W.\ E.\ Ormand,
Phys.\ Rev.\ C {\bf 48}, 1518 (1993); Y.\ Alhassid, D.\ J.\ Dean,
S.\ E.\ Koonin, G.\ H.\ Lang, and W.\ E.\ Ormand, Phys.\ Rev.\
Lett.\, {\bf 72}, 613 (1994).
%
\bibitem{smmc-app} For a recent review, see Y.\ Alhassid, Int.\ J.\ Mod.\ Phys.\
B, {\bf 15}, 1447 (2001).
%
\bibitem{liu}
S.\ Liu, and Y.\ Alhassid, Phys.\ Rev.\ Lett.\ {\bf 87}, 022501
(2001).
%
\bibitem{level}
Y.\ Alhassid, G.\ F.\ Bertsch, and L.\ Fang, Phys.\ Rev.\ C {\bf
68}, 044322 (2003).
%
\bibitem{inertia} Y.\ Alhassid, G.\ F.\ Bertsch, and L.\ Fang,
and S. Liu, Phys.\ Rev.\ C {\bf 72}, 064326 (2005).
%
\bibitem{LG92} E. Y. Loh, Jr. and J. E. Gubernatis, in {\it Electronic
Phase Transitions}, edited by W. Hanke and Y. V. Kopaev (North
Holland, Amsterdam, 1992).
%
\bibitem{vanhoucke}
K.\ van Houcke, S.\ M.\ A.\ Rombouts, L.\ Pollet, Phys. Rev. B
{\bf  73}, 132509 (2006).
%
\bibitem{murthy} G. Murthy, Phys. Rev. B {\bf 70}, 153304 (2004).
%
\bibitem{hstrans}
J.\ Hubbard, Phys.\ Rev.\ Lett. {\bf 3}, 77 (1959); R.\ L.\
Stratonovich, Dokl.\ Akad.\ Nauk.\ S.S.S.R. {\bf 115}, 1097
(1957).
%
\bibitem{ormand} W.E. Ormand, D.J. Dean, C.W. Johnson, G.H. Lang
 and S.E. Koonin, 1994, Phys. Rev. C {\bf  49}, 1422.
%
\bibitem{alhassid00} Y.~Alhassid, Rev. Mod. Phys. {\bf 72}, 895 (2000).
%
\bibitem{ALN99} Y. Alhassid, S. Liu and H. Nakada, Phys. Rev. Lett. {\bf 83}, 4265 (1999).
%
\bibitem{shankar} R. Shankar, Rev. Mod. Phys. {\bf 66}, 129 (1994).
%
\bibitem{halperin}
S.\ D.\ Berger, and B.\ I.\ Halperin, Phys.\ Rev.\ B {\bf 58},
5213 (1998).
%
\bibitem{rombouts} S. Rombouts, D. Van Neck and J. Dukelsky, Phys. Rev. C {\bf
 69}, 061303 (2004).
%
\end{thebibliography}
\end{document}